\documentclass{iau}
\usepackage{graphicx}
\usepackage{hyperref}

\title[Merger origin of Andromeda II] %% give here short title %%
{The major merger origin\\ of the Andromeda II kinematics}

\author[Ebrov\'{a}, {\L}okas, Fouquet, \& del Pino] %% give here short author list %%
{Ivana Ebrov\'{a}, Ewa L. {\L}okas, Sylvain Fouquet, \and Andr\'{e}s del Pino}

\affiliation{Nicolaus Copernicus Astronomical Center, Polish Academy of Sciences,\\ Bartycka 18, 00-716 Warsaw, Poland}

\pubyear{2018}
\volume{S344} %% insert here IAU Symposium No.
\jname{Dwarf Galaxies: From the Deep Universe to the Present}
\editors{K. McQuinn \& S. Stierwalt, eds.}
\begin{document}

\maketitle

\begin{abstract}
Prolate rotation (i.e. rotation around the long axis) has been reported for two Local Group dwarf galaxies:
Andromeda\,II, a dwarf spheroidal satellite of M31, and Phoenix, a transition type dwarf galaxy. The prolate rotation
may be an exceptional indicator of a past major merger between dwarf galaxies. We showed that this type of rotation
cannot be obtained in the tidal stirring scenario, in which the satellite is transformed from disky to spheroidal by
tidal forces of the host galaxy. However, we successfully reproduced the observed Andromeda\,II kinematics in
controlled, self-consistent simulations of mergers between equal-mass disky dwarf galaxies on a radial or close-to-radial
orbit. In simulations including gas dynamics, star formation and ram pressure stripping, we are able to reproduce more
of the observed properties of Andromeda\,II: the unusual rotation, the bimodal star formation history and the spatial
distribution of the two stellar populations, as well as the lack of gas. We support this scenario by demonstrating the
merger origin of prolate rotation in the cosmological context for sufficiently resolved galaxies in the Illustris
large-scale cosmological hydrodynamical simulation.
%% add here a maximum of 10 keywords, to be taken from the file <Keywords.txt>
\end{abstract}

\firstsection % if your document starts with a section,
\begin{figure}[b]
% \vspace*{-2.0 cm}
\begin{center}
\includegraphics[width=\linewidth]{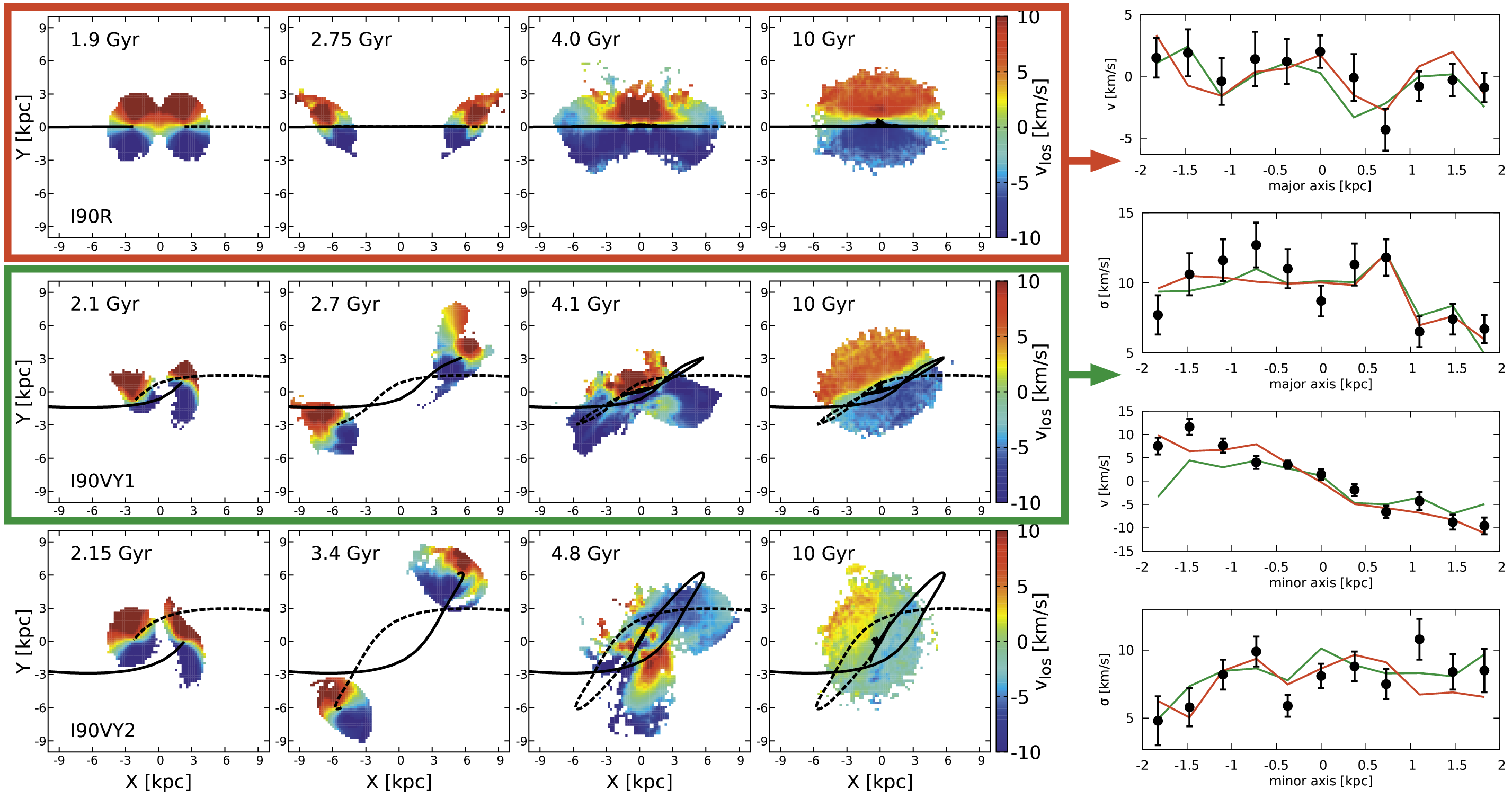}
% \vspace*{-1.0 cm}
\caption{Left: Genesis of prolate rotation in three \textit{N}-body simulations of equal-mass mergers of disky dwarf
galaxies with different initial orbital angular momenta of the colliding galaxies. Black lines indicate the orbits
followed by the dwarfs up to the given time. Each panel shows the area of $20\times 20$\,kpc. Right: Comparison of mean
line-of-sight velocity and velocity dispersion along the axes as observed in And\,II (black points: data from \cite[Ho
et~al. 2012]{ho12}) and produced in the simulations: the radial merger (red; I90R, top row on the left) and the
slightly non-radial merger (green; I90VY1, middle row on the left). Simulated data are computed along the appropriate
line of sight for about 500 stellar particles, mimicking the spatial coverage of the observations. The Figure is based
on Figs.\,3 and 15 from \cite[Ebrov{\'a} \& {\L}okas
(\href{https://doi.org/10.1088/0004-637X/813/1/10}{2015})]{e15andii}. }
\label{fig1}
\end{center}
\end{figure}
%-----------------------------------------

%%%%%%%%%%%%%%%%%%%%%%%%%%%%%
\section{Introduction}

Andromeda\,II (And\,II) dwarf spheroidal galaxy (dSph) is a relatively luminous satellite of M31. It has a stellar mass
of $\sim10^7$\,\(\textup{M}_\odot\), a half-mass radius ~1\,kpc, and ellipticity 0.1--0.2 (e.g., \cite[McConnachie \&
Irwin 2006]{mci06} and \cite[del Pino et~al. 2017]{dp17}). Unlike most dSphs, And\,II contains not only old stars
($>10$\,Gyr) but also significant intermediate-age population. The two stellar populations differ in age,
metallicity, and density profiles (\cite[McConnachie et~al. 2007]{mcc07}). \cite[Ho et~al. (2012)]{ho12} detected
prolate rotation (also known as minor-axis rotation): the rotation around the major axis of the galaxy, i.e. the angle
measured between photometric and kinematic axes is close to 90\,deg. Later, the prolate rotation was also reported in
Phoenix, a transition-type dwarf (i.e., a galaxy that displays intermediate properties between a dwarf irregular and a
dwarf spheroidal; \cite[Kacharov et~al. 2017]{kach17}). And\,II, and possibly Phoenix, could be exceptional examples
of remnants of past major mergers of dwarf galaxies.

A classical scenario known to lead to the formation of dSph galaxies is based on a long-term interaction of the dwarfs
with their host galaxies such as the Milky Way or M31. Initially disky dwarf satellites are transformed to gasless
spheroidals due to tidal forces, dynamical friction and ram-pressure stripping (e.g., \cite[Mayer et~al. 2001]{ma01}).
In \cite[{\L}okas et~al. (2014)]{lo14andii} and \cite[Ebrov{\'a} \& {\L}okas (2015)]{e15andii}, we argued that tidal
interaction is able to remove the initial disk rotation as it transforms to a bar and then a spheroid, but cannot
induce any significant prolate rotation (see also \cite[{\L}okas et~al. 2015]{lok15ts}). Here we review how the
observed properties of And\,II, and prolate rotation in general, can be reproduced via mergers of galaxies.

%-----------------------------------------
\begin{figure}[t]
% \vspace*{-2.0 cm}
\begin{center}
\includegraphics[width=1.0\linewidth]{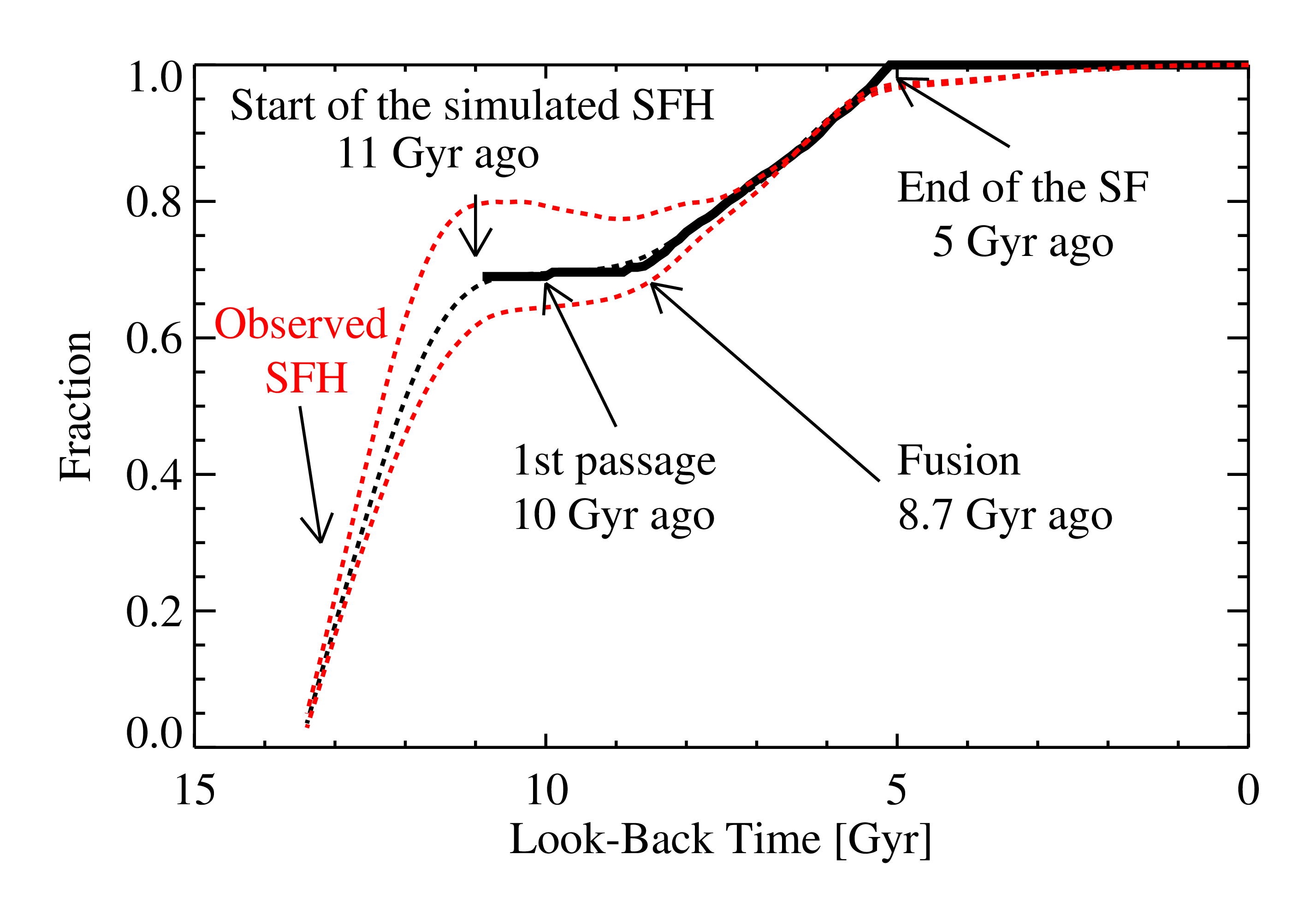}
% \vspace*{-1.0 cm}
\caption{Simulated star formation history of And\,II (thick black line) compared to the observed one (dashed lines;
\cite[del Pino et al. 2017]{dp17}). Red lines show the observational uncertainties. The main stages of the major merger
are indicated. The Figure is adapted from Fig.\,7 in \cite[Fouquet et~al.
(\href{https://doi.org/10.1093/mnras/stw2510}{2017})]{f17andii}. }
\label{fig2}
\end{center}
\end{figure}
%-----------------------------------------

%-----------------------------------------
\begin{figure}[t]
% \vspace*{-2.0 cm}
\begin{center}
\includegraphics[width=1.0\linewidth]{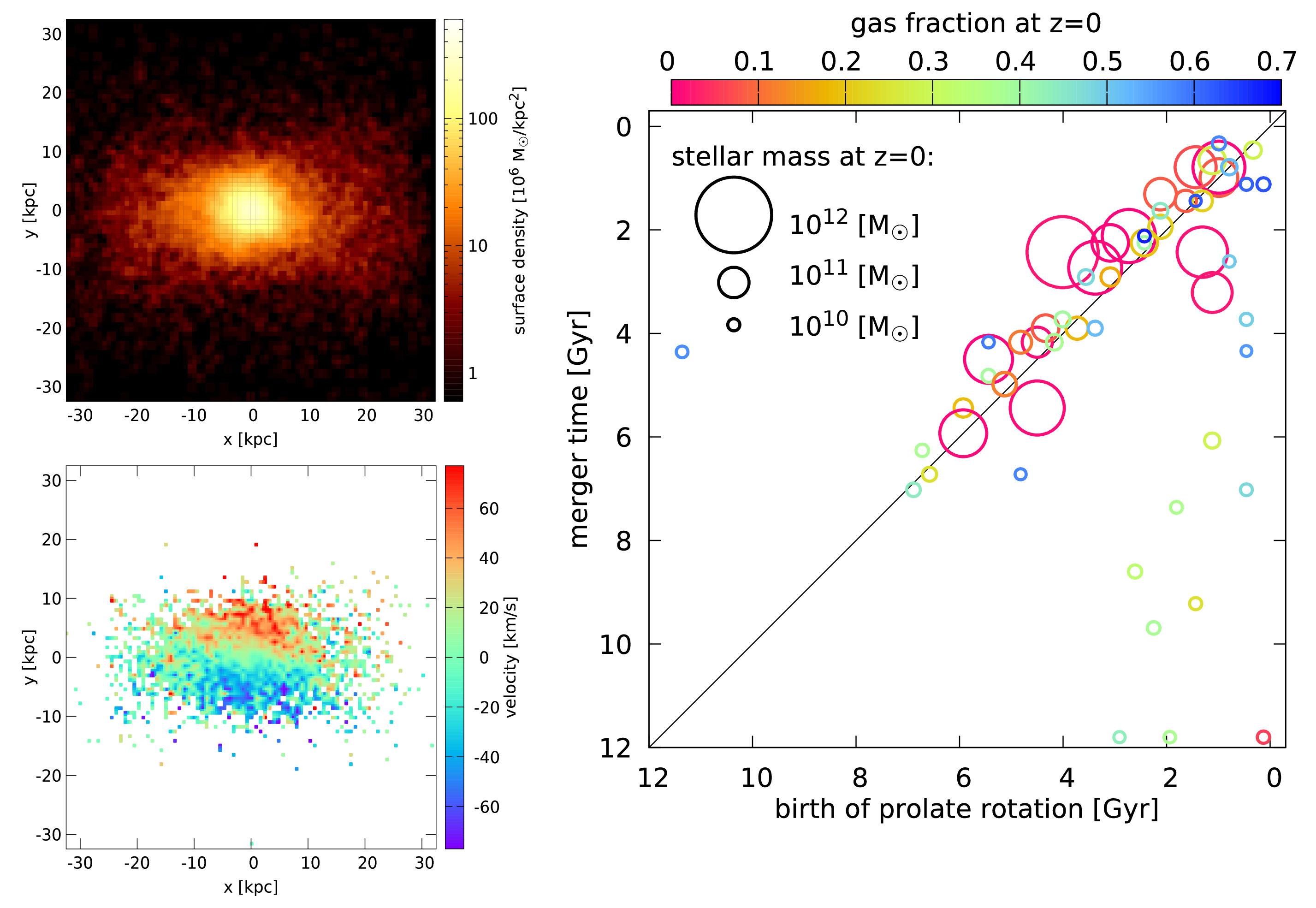}
% \vspace*{-1.0 cm}
\caption{Left: Example of a galaxy with prolate rotation from the final snapshot of the Illustris-1 run. The galaxy has
stellar mass $\sim10^6$\,\(\textup{M}_\odot\) and it maintains prolate rotation since it underwent a major merger
4.17\,Gyr ago. Right: Correlation of the look-back time of the last significant merger (i.e. stellar mass ratio 0.1 or
greater) with the look-back time at which prolate rotation emerged for 59 prolate rotators from Illustris. Circle sizes
are proportional to the stellar mass and colors reflect the gas fraction at the last snapshot. The Figure is based on
Figs.\,1 and~7 from \cite[Ebrov{\'a} \& {\L}okas (\href{https://doi.org/10.3847/1538-4357/aa96ff}{2017})]{el17ill}. }
\label{fig3}
\end{center}
\end{figure}
%-----------------------------------------

%%%%%%%%%%%%%%%%%%%%%%%%%%%%%
\section{The merger origin of prolate rotation}

Using Gadget-2 (\cite[Springel 2005]{sp05}), we performed a series of collisionless self-consistent $N$-body
simulations of dwarf mergers with different initial conditions leading to the formation of a dSph with prolate
rotation. Both merger progenitors had disk mass of $2\times10^7$\,\(\textup{M}_\odot\) ($10^6$ particles) and a
Navarro-Frenk-White dark halo with mass of $10^9$\,\(\textup{M}_\odot\) ($10^6$ particles). In the remnants of the
radial mergers, the amount of rotation is controlled by the inclination of the disks with respect to the collision axis
and all the radial mergers with disk inclinations 60, 90, and 120\,deg reproduce the observed And\,II kinematics. The
non-radial orbits lead to prolate rotation if the orbital angular momentum is initially not much larger than the
intrinsic angular momentum of the progenitor disks. The left part of Fig.\,\ref{fig1} shows snapshots from three
simulations with the same initial disk inclination 90\,deg but with a different initial ratio of the tangential to
radial velocity, from top to bottom: 0, 0.125, and 0.25. The remnants of the most non-radial merger have rotation too
low to account for And\,II but the other two reproduce the observed And\,II kinematics (the right side of
Fig.\,\ref{fig1}) as well as the elliptical shape. The orbital structure of all the simulated merger remnants is
dominated by box orbits in the center and long-axis tubes in the outer parts. For more details see \cite[Ebrov{\'a} \&
{\L}okas (2015)]{e15andii}.

In a follow-up study, we performed a two-stage hydrodynamical simulation including cooling, star formation,
and feedback. An equal-mass radial merger of disky dwarfs with initial gas fraction of 0.7 was followed by a
ram-pressure stripping of the merger remnant in the hot gaseous halo of M31. Such a simulation successfully reproduces
the star-formation history of And\,II, see Fig.\,\ref{fig2}, as well as other observed properties: the magnitude of the
prolate rotation, the mass and shape, the lack of gas, and the spatial distribution of the two stellar populations. For
more details see \cite[Fouquet et~al. (2017)]{f17andii}.

We explored the origin of the prolate rotation using publicly available data from the Illustris large-scale
hydrodynamical cosmological simulation (AREPO moving-mesh code, 106.3\,Mpc box size, $6\times10^9$ initial hydrodynamic
cells and $6\times10^9$ dark matter particles; \cite[Nelson et~al. 2015]{nel15illpub}, \cite[Vogelsberger et~al.
2014]{vog14illpreintro}, \cite[Rodriguez-Gomez et~al. 2015]{rg15illmer}). From the last snapshot, we selected 59 well
resolved galaxies with well-established prolate rotation (see the left part of Fig.\,\ref{fig3} for one example). These
prolate rotators are more abundant among massive galaxies just as reported for the observed galaxies (\cite[Tsatsi
et~al. 2017]{tsa17}, \cite[Krajnovi{\'c} et~al. 2018]{kr18}). The birth of prolate rotation strongly correlates with
the last significant merger (Fig.\,\ref{fig3} on the right). The mergers leading to prolate rotation have a wide range
of initial conditions but they are slightly more radial and have higher mass ratios and more recent merger times than
mergers of the comparison sample. About half of Illustris prolate rotators are created in gas-rich mergers. For more
details see \cite[Ebrov{\'a} \& {\L}okas (2017)]{el17ill}.

%%%%%%%%%%%%%%%%%%%%%%%%%%%%%
\section{Conclusions}

We demonstrated that prolate rotation, detected in And\,II and Phoenix dwarf galaxies, which cannot be reproduced in
the scenario of disky dwarfs interacting with the host galaxy, can be generated in a major merger of two disky dwarfs on
a radial or close-to-radial orbit. The observed And\,II kinematics, the star formation history, the spatial
distribution of the two stellar populations, and the lack of gas is best explained in a simulation of a gas-rich merger
of two disky dwarfs about 9\,Gyr ago followed by ram-pressure stripping caused by a close interaction with M31 about
5\,Gyr ago. We support this scenario by demonstrating the merger origin of prolate rotation in the cosmological context
in the Illustris simulation.

%%%%%%%%%%%%%%%%%%%%%%%%%%%%%
\acknowledgments
This research was supported by the Polish National Science Centre under grants 2013/10/A/ST9/00023 and
2017/26/D/ST9/00449.

%%%%%%%%%%%%%%%%%%%%%%%%%%%%%

\end{document}